\documentclass[preprint,eqsecnum,aps]{revtex4}     
\usepackage{epstopdf}
\usepackage{graphicx}
\usepackage{amsmath}
\usepackage{epsfig}
\def\figdir{erik_LZS/figures}
\def\figdir2{erik_LZS}
\def\beq{\begin{equation}}
\def\eeq{\end{equation}}
\def\bea{\begin{eqnarray}}
\def\eea{\end{eqnarray}}
\def\nnu{\nonumber}
\def\tst{\textstyle}
\def\dst{\displaystyle}

\def\fno#1{Fig.~\ref{#1}}

\def\eno#1{Eq.~(\ref{#1})}
\def\Eno#1{Equation (\ref{#1})}
\def\etwo#1#2{Eqs.~(\ref{#1}) and (\ref{#2})}

\def\Sno#1{Sec.~\ref{#1}}

\def\by{\over}
\def\gtwid{\mathrel{\raise.3ex\hbox{$>$\kern-.75em\lower1ex\hbox{$\sim$}}}}
\def\ltwid{\mathrel{\raise.3ex\hbox{$<$\kern-.75em\lower1ex\hbox{$\sim$}}}}

\def\al{\alpha}

\def\dta{\delta}
\def\eps{\epsilon}
\def\veps{\varepsilon}

\def\sig{\sigma}

\def\Gam{\Gamma}
\def\Dta{\Delta}

\def\Tta{\Theta}

\def\apx{\approx}
\def\ptl{\partial}
\def\hf{\frac{1}{2}}
\def\tshf{{\tst\hf}}

\def\tofro{\leftrightarrow}

\def\inmipi{\int_{-\infty}^{\infty}}

\def\dint{\int\!\!\!\int}
\def\grad{\nabla}

\def\part#1#2{\frac{\ptl#1}{\ptl#2}}

\def\lf{\left}
\def\rt{\right}

\def\ham{{\mathcal{H}}}
\def\ket#1{|#1\rangle}

\def\bB{{\bf B}}
\def\bH{{\bf H}}

\def\bM{{\bf M}}
\def\bS{{\bf S}}

\def\nhat{{\bf{\hat n}}}

\def\itx{{\it x\ }}

\def\itxy{{\it xy\ }}

\def\Fe8{Fe$_8$}
\def\Mn12{Mn$_{12}$}

\def\up{\uparrow}
\def\dn{\downarrow}
\def\kup{\ket{\!\up\,}}
\def\kdn{\ket{\!\dn\,}}

\def\fsi{f_{\sig}}
\def\fsb{f_{\bar\sig}}
\def\gssb{\Gam_{\sig{\bar\sig}}}
\def\gsbs{\Gam_{{\bar\sig}\sig}}
\def\gspp{\Gam_{{\bar\sig}'\sig'}(\eps')}
\def\gepe{g\bigl(\tshf(\eps' - \eps)\sig' \bigr)}
\def\gpepe{g_R\bigl(\tshf(\eps' - \eps)\sig' \bigr)}
\def\geppe{g\bigl(\tshf(\eps'' - \eps)\sig' \bigr)}
\def\geep{g\bigl(\tshf(\eps - \eps')\sig' \bigr)}
\def\geepp{g\bigl(\tshf(\eps - \eps'')\sig' \bigr)}
\def\tsbs{T_{{\bar\sig}\sig}}
\def\tspp{T_{{\bar\sig}'\sig'}}

\begin{document}


\title{Low-Temperature Magnetization Dynamics of Magnetic Molecular Solids in a Swept Field}
\author{Erik Lenferink}
\author{Avinash Vijayaraghavan}
\author{Anupam Garg}
\email[e-mail address: ]{agarg@northwestern.edu}
\affiliation{Department of Physics and Astronomy, Northwestern University,
Evanston, Illinois 60208}

\date{\today}

\begin{abstract}
The swept-field experiments on magnetic molecular solids such as \Fe8 are studied using Monte Carlo simulations, and
a kinetic equation developed to understand collective magnetization phenomena in such solids, where the collective aspects
arise from dipole-dipole interactions between different molecules. Because of these interactions, the classic
Landau-Zener-Stuckelberg theory proves inadequate, as does another widely used model constructed by Kayanuma.
It is found that the simulations provide a quantitatively accurate account of the experiments. The kinetic equation
provides a similarly accurate account except at very low sweep velocities, where it fails modestly. This failure is
attributed to the neglect of short-range correlations between the dipolar magnetic fields seen by the molecular spins.
The simulations and the kinetic equation both provide a good understanding of the distribution of these dipolar fields,
although analytic expressions for the final magnetization remain elusive.
\end{abstract}

\maketitle

\newpage
\section{Introduction}
\label{intro}

As a prototype of magnetic molecular solids, the one generally known as \Fe8 has drawn great interest. This material
consists of molecules of [Fe$_8$O$_2$(OH)$_{12}$(tacn)$_6$]Br$_8$(H$_2$O)$_9$, in which the Fe(III) ions of one
molecule are well separated from those of a neighboring molecule. At low temperatures, each molecule has a spin
$S = 10$, a corresponding all-spin magnetic moment of magnitude $g\mu_B S$ with $g \simeq 2$, and an
Ising-like anisotropy which translates into an energy barrier of $~22\,$K~\cite{fe8props,gat06}. Among the many 
experimental investigations of ths molecule, the swept-field experiment of Wernsdorfer and Sessoli~\cite{wer99} is one
of the most revealing. A partial and simplified description of this experiment is as follows. At low temperatures
($T \ltwid 100\,$mK) they first saturate the magnetization of the sample by applying an external magnetic field $H_z$
along the Ising or $z$ axis, and also apply a magnetic field $H_x$ along the hard magnetic axis transverse to
the easy axis. They then sweep $H_z$ so as to reverse the magnetization, and measure the rate at which the
spins reverse. This rate turns out to be an oscillatory function of $H_x$, even though the energy barrier and
the angle between the two energy minimizing orientations of the spin are both monotonically decreasing functions
of $H_x$. Although surprising at first, the oscillations are now well understood. The simplest explanantion for
the magnetization reversal is that the sweeping of $H_z$ induces Landau-Zener-St\"uckelberg (LZS)
transitions~\cite{lan77,kay84} between the two lowest states on opposite sides of the energy barrier at a rate
proportional to $\Dta^2$, where $\Dta$ is the tunnel splitting between these states. The rate oscillates because
$\Dta$ does so, and $\Dta$ in turn oscillates because there are two tunneling spin trajectories that
interfere with a phase that varies with $H_x$~\cite{los92,vdel92,gar93}.

\Fe8 is just one of $\sim10^3$ magnetic molecular solids that are now known, and which have been the object of much study
over the last two decades. Their main characteristics are that the spin of one molecule is large at low temperatures,
and to a good first approximation, one may treat the spins on different molecules as non-interacting.
(Acomprehensive and authoritative review of the entire field is contained in Ref.~\cite{gat06}. Shorter reviews
may be found in Refs.~\cite{can99,vil00,fri10}.) These solids are often known as single-molecule-magnets (SMM's),
because many phenomena may be understood, at least qualitatively, in terms of the properties of the total ground
state spin of a single molecule in a suitable crystal field, via an effective spin Hamiltonian that contains
anisotropy terms reflecting the overall symmetry of the molecule and its local environment. The oscillatory tunnel
splitting mentioned above is one such phenomenon. Another is that the hysteresis loops are sharply stepped, where
the steps coincide with crossings of energy levels on opposite sides of the energy barrier~\cite{fri96,tho96}.
This single molecule behavior has, unsurprisingly, led to suggestions and proposals for using these materials
in devices~\cite{bog07}, but they are a new class of magnetic materials and worthy of study in their own right
for the novel phenomena they display.

A large variety of experimental tools has been employed to study the low temperature magnetization dynamics of such
solids. Of these, the swept-field or Landau-Zener-St\"uckelberg (LZS) protocol has proven to be one of the most
fruitful. When the sweep is sufficiently rapid, the accompanying change in the magnetization can be interpreted
in terms of LZS transitions as already mentioned, and thereby provides a measurement of the tunneling amplitude
between energy levels on opposite sides of a barrier. Tunnel splittings measured by this technique are as low
as $10^{-8}$\,K in temperature units. Such low splittings are beyond the reach of any other method. When the sweep
rate is slow, on the other hand, the interpretation is not clear-cut. It is essential to consider the dipole-dipole
interactions between different molecules, and the transition is influenced by the collective dynamics of all the
spins. Indeed, in this case, the SMM designation falls short, and a more intricate analysis is called for.

Collective, dipole-coupled dynamics of the spins are also seen in several magnetization-relaxation type
experiments~\cite{san97,ohm98,wer99b,tho99,wer00,wer00b,tup02}. Theoretical discussions and Monte-Carlo simulations
of these experiments have been provided by~\cite{gat06,pro98ab,cuc99,fer0304,vij12}, and many aspects of the
experiments are understood. The same is not true of the swept-field experiments, and we are aware of only a few
previous investigations that pertain to the collective dynamical aspects~\cite{liu02,gara03,fu04}. We discuss these
below. Theories that include the dipole-dipole interactions in a purely static way~\cite{kec07}, or via a dynamic
mean-field~\cite{avag09} cannot explain all the experimental behavior, especially that at low sweep rates.
It is important to have a more complete theory since experimentalists often interpret data in terms of the original
LZS analysis~\cite{ram08,wer08,ram08b,wer08b}. In light of the fact that the spins in molecular magnets are subject
to strong environmental influences, especially the dipolar couplings to other molecular spins, and that the
LZS theory is based on fully quantum mechanically coherent time evolution, it is not at all clear that the LZS
description is even applicable a priori. Indeed, as explained in Ref.~\cite{avag09}, it is a fortunate fact that it
can be used even at high sweep velocities. Wernsdorfer has argued~\cite{wer08}, correctly in our view, that
the LZS formula should not be used without first verifying that one is in the fast sweep limit, and that it can
in general only yield a lower bound on the tunnel splitting. While the use of this formula may provide qualitative
proof of the presence of a geometrical phase in the tunneling spectrum~\cite{gar93}, one is on shaky ground if
one uses the extracted tunnel splittings to build detailed models of intramolecular magnetic interactions. It seems
to us that a recent debate about these matters~\cite{ram08,wer08,ram08b,wer08b,ram09,wer09} is caused by an
incomplete understanding of how molecular magnets respond to a swept magnetic field.

It is the purpose of this paper to provide a more complete analysis of the swept-field protocol including the
collective behavior of all the spins. We study the coupled system of molecular spins by Monte-Carlo simulations
and by developing and solving a kinetic equation for the joint probability distribution of the spin (up or down)
of a molecule and the local magnetic field seen by that spin. We compare the results of these two approaches with
each other, and with experiments. We find that the Monte Carlo simulation agrees with the experimental results over
almost the entire range of sweep velocities, giving us confidence that the physical picture underlying the
simulations is correct. The kinetic equation agrees with the simulations up to moderately low sweep velocities,
but fails at still slower velocities, although the failure is not as severe as for previous theoretical
treatments~\cite{avag09}. In all cases, we find that a model due to Kayanuma~\cite{kay84} of a spin strongly coupled
to a bath does a better job of explaining the physics than the LZS theory, but it is never superior to the kinetic
equations or the Monte Carlo simulations. The main advantage of the Kayanuma model is that it yields a simple
formula for the magnetization reversal in terms of the tunneling matrix element, whereas we are unable to write
a similar formula for the results of our kinetic equation or Monte Carlo simulations.

We note here that a kinetic equation was written down in \cite{pro98ab} using a formal approach wherein the collision
or interaction term between pairs of spins was expressed in terms of the two-site distribution of spin and local
magnetic field. It is thus like the first of the equations of the BBGKY hierarchy, which are exact, but which
cannot be solved without truncating the hierachy. The simplest truncation is at the level of the single-site
distribution itself, and amounts to assuming that the two-site distribution factorizes. However, rather than obtain
the kinetic equation in this formal way, we derive the equation by considering the various interaction processes
which cause the (one-site) distribution to change. This approach gives greater physical insight into the
approximations made. The single-site approximation breaks down at ultra slow sweeps as we explain, leading to the
divergence from the simulations mentioned above. Including two-site correlations would lead to an very much more
complex numerical problem, however, and experience with other problems suggests that if two-site correlations are
truly important, then one is faced with a humdinger, as three-site, four-site, and higher-order multi-site
correlations are also likely to be important.

A few brief comments on Ref.~\cite{liu02}, a very interesting study of exactly the same problem we study here, may
also be in order. These authors write an equation for the rate of change of magnetization as the product of two
factors. The first is essentially the reversible magnetization, and is obtained by multiplying the net magnetization
with the distibution of dipole fields evaluated at zero field value. In this part, the approach has much the
same philosophy as ours. The second factor is the probability for an individual spin to flip. This factor is
evaluated by considering all the spins which pass through the LZS crossing at the same time as a many-body problem,
whose quantum mechanical evolution is fully coherent. This many-body problem is then reduced to a one-body problem
via a mean field approximation, but the fact that the transition is coherent is still retained. In our view, this is
unlikely to be the case in the experiments, and the nuclear spins are an obvious mechanism that render the
tunneling incoherent~\cite{pro96,avag09}. It may be possible to study a system with no nuclear moments whatsover,
but even then, the notion of coupled, fully coherent, simultaneous LZS transitions of many spins seems to us an
extremely difficult one to realize in practice. It is a general rule that low energy environmental degrees of freedom
are especially damaging to macroscopic quantum coherence~\cite{sixman}. Given the extremely small energy scale of
the tunnel splittings, almost any low energy environmental degree of freedom that one normally ignores at the
temperatures of the experiments (such as phonons) can be expected to spoil the coherence. Thus, even though
Ref.~\cite{liu02} is able to fit the  experimental data well, the basic premises of the theory seem implausible to us.
Refs.~\cite{gara03,fu04} also consider only fully quantum mechanically coherent coupled LZS transitions, and are not
germane to the real systems.

The rest of the paper is organized as follows. In \Sno{background}, we provide background information on \Fe8, the
LZS and Kayanuma models, and on a theoretical model for SMM's that incorporates the influence of the
environment~\cite{avag09,pro96}. We also describe how this model differs from our earlier rate equation
approach~\cite{vij12}. In \Sno{thefullmonty}, we discuss our Monte Carlo protocol, and the results of our simulations.
The kinetic equation is discussed in \Sno{kin_eqn}. We first derive this equation, and then discuss how various
partial sums of the collision term in this equation may be physically interpreted. We also discuss how we integrate
this equation numerically, and the results of this integration. Section \ref{conc} contains a brief summary of our
conclusions.

\section{Background Information and Models}
\label{background}

\subsection{Independent and Coherent Spin Approximation}
\label{indy_spin}

Let us first assume (counterfactually) that the interactions between different molecular spins may be ignored.
The dynamics of the spin of a single molecule would then be governed by an anisotropy Hamiltonian of the form
\beq
\ham = -k_2 S_z^2 + \ham_{\perp} - g\mu_B \bS\cdot\bH,
\eeq
where $\bS = (S_x, S_y, S_z)$ is the total spin of the molecule, the first term in $\ham$ is the leading anisotropy,
$g$ is the {\it g\/}-factor, $\mu_B$ is the Bohr magneton, $\bH$ is the external magnetic field, and $\ham_{\perp}$
is a term in the transverse spin components that is off-diagonal in the $S_z$ basis and reflects the intrinsic
higher-order anisotropies of the molecule. For this reason it is time-reversal invariant, and so contains only even
powers of the spin components. In \Fe8, for example,
\beq
\ham_{\perp} = (k_1 - k_2) S_x^2 - C(S_+^4 + S_-^4). \label{Hperp_fe8}
\eeq
The various parameters in $\ham$ are well known for the most highly studied molecules. For \Fe8, $S=10$, $g\simeq 2$,
$k_1 \simeq 0.33\,$K, $k_2 \simeq 0.22\,$K, and $C \simeq 29\,\mu$K. In \Mn12, by contrast, the anisotropy is
tetragonal based on the symmetry of the molecule, but there are believed to be biaxial terms of the same type as
in \eno{Hperp_fe8} arising from chemical disorder, variable waters of crystallization, variant chemical species,
etc., and there are also additional longitudinal terms such as $B S_z^4$. These details are largely immaterial for
this paper, since we are interested in low temperatures only. We may therefore focus on the lowest two energy levels,
$m = \pm S$, and presuppose that there is an amplitude per unit time, $-i\Dta/2\hbar$, to tunnel between these levels.
We also assume that the effects of the transverse fields $H_x$ and $H_y$ have been incorporated in $\Dta$, and only
the longitudinal field, $H_z$, is not. It is this field that is swept.

Under these conditions, each molecular spin may be described as a two-level system governed by an effective
single-spin Hamiltonian,
\beq
\ham_{\rm eff}
   = \frac{1}{2}
        \begin{pmatrix}
                \eps(t) & \Dta \\
                \Dta & -\eps(t) \\
        \end{pmatrix},  \label{ham2by2}
\eeq
where $\eps(t)$ is the energy of the $m=S$ state relative to that of the $m=-S$ state, given by
\beq
\eps(t) = 2Sg \mu_B H_z(t).  \label{eps_vs_H}
\eeq
We shall refer to $\eps$ as the {\it bias\/} on the spin~\cite{bias}. In writing \eno{eps_vs_H}, we have defined
the zero of $H_z$ so that the levels $\pm S$ are degenerate at $H_z = 0$. In the laboratory, a nonzero offset field
may be required to cancel demagnetizing fields and bring this degeneracy about; $H_z$ is supposed measured from
this offset. We will interchangeably refer to the two states either as $m = \pm S$ states, or as pseudospin-1/2 states
$\kup \equiv \ket{m= +S}$ and $\kdn \equiv \ket{m = -S}$ states, or as ``up" and ``down" states.

We now suppose that the longitudinal field is swept at a steady rate so that
\beq
\eps(t) = \dot\eps t,
\eeq
and the spin is in the lower energy state, $\kup$ as $t \to -\infty$. Then, the probability that spin will flip into the
$\kdn$ state as $t\to \infty$ is given by the classic LZS formula
\beq
  P_{\rm LZS} = 1 - \exp\bigl(-\pi\Dta^2/2 |\dot\eps| \bigr).
\eeq
The limits of fast sweep, $\dot\eps \gg \Dta^2$, and slow sweep, $\dot\eps \ll \Dta^2$, are worth noting:
\beq
P_{\rm LZS}
  \apx \begin{cases}
          \dst{\frac{\pi\Dta^2}{2 |\dot\eps|}},  & \dot\eps \gg \Dta^2, \cr
          \noalign{\vskip4pt}
          1,                                     & \dot\eps \ll \Dta^2. \cr
       \end{cases}
\eeq

\subsection{The Kayanuma model}
\label{kayanuma}

As mentioned in \Sno{intro}, the LZS model is inadequate to describe the actual experiments. An alternative
single-spin model is that of Kayanuma~\cite{kay84} wherein the bias field $\eps(t)$ has added to it a fluctuating
part $\eta(t)$, which is taken as a Gaussian random process. In the limit where this process has very large amplitude
and is delta-function correlated in time (white noise), corresponding to a very rapidly fluctuating bias, the
probability that the spin will end up in the state $\kdn$ having started in $\kup$ is found to be
\beq
  P_{\rm K} = \hf \Bigl(1 - \exp\bigl(-\pi\Dta^2/ |\dot\eps| \bigr)\Bigr). \label{p_kaya}
\eeq
The fast and low sweep limits of the reversal probability are now given by
\beq
P_{\rm K}
  \apx \begin{cases}
          \dst{\frac{\pi\Dta^2}{2 |\dot\eps|}},    & \dot\eps \gg \Dta^2, \cr
          \noalign{\vskip4pt}
          \dst{\hf},                               & \dot\eps \ll \Dta^2. \cr
       \end{cases}
\eeq
This is identical to the LZS formula for fast sweeps, but very different for slow ones. Whereas the LZS process
describes coherent adiabatic reversal of the spin, the Kayanuma process describes a spin which is able to make
many transitions between the up and down states and is therefore completely randomized at $t = \infty$. We may think
of the fluctuating field $\eta(t)$ as a qualitative way of describing the dipolar field of other spins, which are
also undergoing transitions between up and down states.

It was shown in Ref.~\cite{avag09} that we also obtain Kayanuma's answer (\ref{p_kaya}) if we assume that the
external field is swept uniformly, and each spin flips between the up and down states independently of the others
but with a bias-dependent rate. That is, if the probability for a spin to be in the up state at time $t$ is denoted
$p_{\up}(t)$, we take
\beq
{\dot p}_{\up} = \Gam[\eps(t)]\,(1 - 2p_{\up}). \label{rate_ind_eps}
\eeq
\Eno{p_kaya} follows if $\eps(t) = {\dot \eps}t$ and
\beq
\inmipi \Gam(\eps)\, d\eps = \frac{\pi}{2}\Dta^2. \label{int_Gam}
\eeq
Since this requirement on $\Gam(\eps)$ entails only the tunneling amplitude $\Dta$,
the details of the physical decoherence mechanism that justifies writing down a rate equation in the first place
are irrelevant. In Ref.~\cite{avag09}, these mechanisms involved the environments of nuclear spins and other
molecular spins. The bias-dependent flip rate that we employ in this paper [see \eno{gamma1}] is a special case of
that in \cite{avag09}, and satisfies \eno{int_Gam}.

The Kayanuma model is attractive because it is simple and economical. It is a much better model for the reversal
than the LZS process. It is plausible that for a large collection of molecular spins where each one may flip at
different times in the sweep cycle because the offset field is not the same for all spins, the field seen by any
one spin has a stochastic character. However, it is clearly an approximation to assume that the time-dependence
of the bias on any one spin is uncorrelated with the configuration of the other spins. As shown in \cite{avag09}, it
is not even enough to take account of the spin configuration in a mean-fieldy way through the spatially averaged
demagnetization field. Therefore, we must consider the interaction between the spins explicitly.

\subsection{Environmentally influenced interacting spins}
\label{glauber_spins}

In reality, the spins are neither isolated nor noninteracting as noted before. As discussed in \cite{avag09,pro96},
there are two types of interactions to consider. First, each molecular spin is coupled to nuclear spins in its
vicinity. These spins have the effect that the tunneling between the $m = \pm S$ states becomes fully incoherent for
typical molecular magnets. The $\kup \tofro \kdn$ probability for the $i$th spin is given by 
\beq
p_{{\rm{flip}},i} = \Gam_i\, dt,
\eeq
where
\beq
\Gam_i \equiv \Gam(\eps_i)
   = {\sqrt{2\pi} \by 4} {\Dta^2 \by W}
          \exp \Bigl(-{\eps_i^2\by 2W^2} \Bigr).  \label{gamma1}
\eeq
Here, $W \simeq 10 E_{dn}$ with $E_{dn}$ being the dipole-dipole interaction energy between the molecular spin and
the nearby nuclear spins, and $\eps_i$ is the bias on site $i$~\cite{bias}. For \Fe8, $E_{dn} \sim 1$~mK, and
$\eps_i \sim 0.1$~K in temperature units (see below).

Second, each molecular spin is coupled to all the other molecular spins via the dipole interaction. At first sight,
this leads to a fully many-body quantum mechanical problem. Since the nuclear spins already render the molecular
spins slow and incoherent, however, the quantum mechanical back action of the latter upon each other has no effect
whatsoever, and may be ignored. Hence, the field of one molecular spin on another may be taken as a c-number. If we
define an Ising spin variable $\sig_i$ on every site $i$, such that $\sig_i = \pm 1$ corresponds to $m_i = \pm S$,
the dipolar part of $\eps_i$ may be written as
\bea
\eps_{i,\rm{dip}} &=& \sum_{j\ne i} K_{ij} \sig_j, \label{Ebias} \\
K_{ij}            &=& 2 {E_{dm} a^3 \by r^3_{ij}}
                      \lf(1-3{z^2_{ij} \by r^2_{ij}} \rt). \label{Kij}
\eea
Here, $E_{dm}$ is the interaction energy scale between neighboring molecular spins, and $a$ is their separation.
Further, $r_{ij}$ is the distance between spins $i$ and $j$, and $z_{ij}$ is the difference between their
{\it z\/} coordinates. We estimate the dipolar field in \Fe8 as $\sim 100$~Oe, which implies the energy
scale $\eps_i \sim E_{dm} \sim 0.1$~K quoted above.

The net result of these two couplings is as follows. The ratio $\eps_i/W$ is large for most molecules most of the
time, so these spins are essentially frozen. A spin is unfrozen only when the field it sees is essentially zero
(more precisely of order $W$ or less). We refer to this range of bias values as the {\it reversibility region\/}.
In magnetization relaxation experiments, the combination of a static and narrow reversibility region leads to very
slow and non-exponential time decay, which many workers have investigated
theoretically~\cite{pro98ab,cuc99,fer0304,vij12}. If the external field is swept, however, the odds of a spin being
able to flip are greatly increased. Particularly if the sweep rate is low, the complete dynamics can be very rich
and complex, as we now discuss.

Let us focus on one spin as the bias is being increased. Since the dipole energy is so large compared to all other
energies in the problem, the flip of even a far away spin can change the bias on the first or central spin from a
big negative value (relative to $W$) to a big positive one. The central spin is then shunted past the $\eps = 0$
region and does not flip. This gives a qualitative explanation of why the net magnetization reversal can be less than
that in Kayanuma's model. However, the central spin can be returned to the region of negative bias if another spin
not too far away were to flip subsequently. Thus, the actual amount of spin reversal is hard to calculate
analytically. We therefore turn to Monte Carlo simulations of the LZS protocol.

\section{Monte Carlo Simulations}
\label{thefullmonty}

The physical problem we have described in \Sno{glauber_spins} is characterized by four parameters: the molecular spin
dipole-dipole energy $E_{dm}$, the reversible region width $W$, the external bias sweep rate $\dot\eps_a$, and the
tunnel splitting $\Dta$. For purposes of analysis and Monte Carlo simulation, however, these parameters can be
reduced to just two dimensionless variables: the scaled reversible region width $w$ and the scaled sweep rate $v$,
given as
\bea
w &=& \frac{W}{E_{dm}}, \\
v &=& \frac{|\dot\eps_a|}{\Dta^2}.
\eea
All bias energies are measured in units of $E_{dm}$. We use the symbol $\veps$ (note the different font) for the
dimensionless biases, both applied and dipolar. It is not always optimal to frame the discussion in terms of
dimensionless variables, and we shall switch back and forth between dimensionless and dimensionful descriptions
as needed for clarity.

\subsection{Simulation protocol}
\label{how_we_simulate}

Our Monte Carlo simulation models a finite system of $N$ spins, labeled $\sig_i$, each of which can be in either the
up or the down state ($\sig_i=\pm1$). The bias $\eps_i$ at site $i$ is the sum of the externally applied bias
field $\eps_a(t)$ and the dipole field $\eps_{i,{\rm dip}}$:
\beq
\eps_i(t) = \eps_a(t) + \eps_{i,{\rm dip}}(t).
\eeq
We have explicitly shown that the dipole field is time dependent since the spin state of the system is time dependent.
The spins are taken to lie on a cubic lattice with lattice constant $a$, and the sample is taken to be spherical so
that the initial bias distribution is approximately a delta function centered at the value $\eps_a(0)$. For the
majority of the simulations, the number of spins, $N$, is chosen to be 7{,}153, corresponding to 25 spins on the
sphere's diameter, and $\eps_a(t)$ is taken to vary linearly in time,
\beq
\eps_a(t) = {\dot\eps}_a t.
\eeq
Though the number $N$ is significantly less than that used for previous simulations of relaxation~\cite{vij12}, we
found that the results converged even for $N \gtwid 1,419$ (diameter $\ge 15$ spins). In simulations of relaxation,
large values of $N$ were necessary to ensure that the reversible region was never depleted entirely. In simulations
of LZS sweeps, the reversible region moves across the bias distribution, so depletion is a much lesser concern.

In each Monte Carlo timestep (which we arbitrarily denote by $dt$), the spin flip probabilities are calculated for
every spin in the system from the rate function $\Gam(\eps)$. To save on computation time, for many of our
simulations, we used a simplification of \eno{gamma1} that is only nonzero for $|\eps| \leq W$:
\beq
	\Gam'(\eps) = \frac{\pi\Dta^2}{4W}\Theta(W-|\eps|).           \label{gamma2}
\eeq
(The constant multiplying $\Tta(W - |\eps|)$ is chosen so that the integrals
of $\Gam(\eps)$ and $\Gam'(\eps)$ over all $\eps$ are identical, and \eno{int_Gam} holds with $\Gam'$.)
In practice, this rate function gives very nearly the same results as \eno{gamma1} and allows a simpler view of the
spin flip process: the only spins that are able to flip are those strictly within the reversible region, a window of
width $2W$ around zero. After the spins have flipped, the change in bias at every lattice site is calculated and the
simulation moves to the next timestep.

Over the course of one run of the simulation, the external bias is swept from $-25 E_{dm}$ to $25 E_{dm}$. If
$d\eps_a$ is the change in the bias in one time step, the probability for a spin in the reversible region to flip in
that timestep is
\bea
  p_{\rm flip} &=& \frac{\pi\Dta^2}{4W} dt \nnu \\
               &=& \frac{\pi}{4} \frac{E_{dm}}{W} \frac{\Dta^2}{E_{dm}} \frac{d\eps_a}{|\dot\eps_a|} \nnu \\
               &=& \frac{\pi}{4wv}d\veps_a.
\eea
Naturally, care must be taken to ensure that $p_{\rm flip} \ll 1$. In practice, we required that it did not exceed
$0.1$. We also want to ensure that we do not skip past the reversible region in just one time step, so $d\veps_a$ is
also chosen not to exceed $0.1 w$. This becomes computationally intensive for slow sweeps, and for $v =0.1$, for
example, we are only able to study values of $w$ exceeding $0.01$.

The quantities that we record during the simulation are the magnetization per site,
\beq
M = \frac{1}{N}\sum_i \sig_i,
\eeq
and the bias field distributions, $f_{\pm}(\eps)$, defined so that $f_{\pm}(\eps) \, d\eps$ is the fraction of sites
with spin $\sig = +1$ or $-1$ and a bias field in the range $\eps$ to $\eps + d\eps$. We are especially interested
in the final magnetization, $M_f$.

We have already touched on the feasibility of performing the simulations with spheres of as few as 15 spins on the
diameter. However, to test the stability of the answers with system size, we have performed some simulations with
as many as 55 spins on the diameter. Further, in some simulations, we have employed a triangular wave for $\eps(t)$
as in some of the experiments~\cite{wer99,wer00,wer00b}. The excursion of the bias is again $\pm 25 E_{dm}$. In
these cases, ${\dot\eps}$ refers to the value of $|d\eps/dt|$ on any leg of the wave. Finally, a few simulations are
done using the full Gaussian rate function (\ref{gamma1}). We shall mention these exceptional cases as they merit.

\subsection{Simulation Results}

\begin{figure}[h!]
\includegraphics[scale=.8]{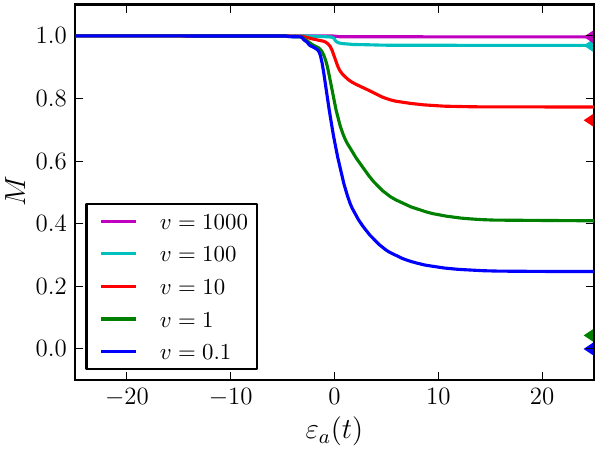}
\caption{(Color online) Monte Carlo results for the magnetization of the system as a function of the (scaled)
externally applied bias as it is swept from negative to positive values for various values of $v$ and $w = 0.05$. Since
$\eps_a$ varies linearly with time, this is essentially a plot of $M$ vs{.} $t$. Each curve is an average of 100 runs
for a system of $N=7{,}153$ spins using the modified transition rate (\ref{gamma2}). The triangles on the right edge
of the plot show $M_{f,{\rm K}}$, the final magnetization for each value of $v$ as per Kayanuma's model. The tiny drop
just before $\eps = 0$ is due to surface spins.}
\label{m-bias}
\end{figure}
In \fno{m-bias} we show a plot of the time dependence of the magnetization for various values of $v$ and $w = 0.5$.
Readers will note that irrespective of $v$, the major change in $M$ starts when $\veps_a$ reaches a value
close to zero, and there is little change once $\veps_a$ exceeds $\sim 5$. Readers will also note that there
appears to be a small drop in $M$ before the main one. The reason is that in the initial state, when all the spins
are up, the bias field is zero at almost every site. However, there are always some spins with nonzero bias
on the surface of the sample, though their fraction becomes smaller as $N$ increases. Thus the initial bias
distribution is not quite a delta function centered at zero bias, and the surface spins start to flip before the
external bias reaches $-w$, and this is responsible for the small dip.

\begin{figure}[h!]
\centering
\includegraphics[scale=.9]{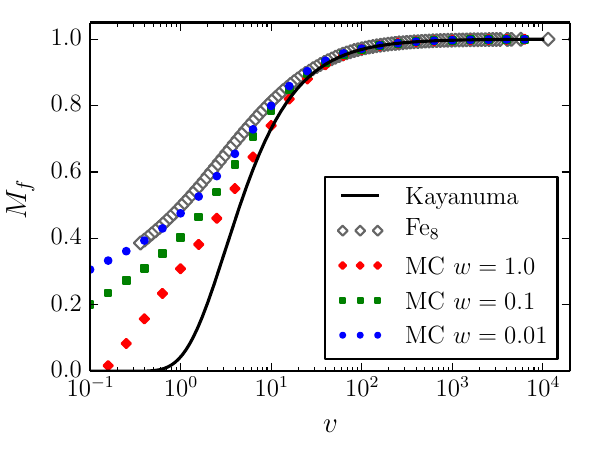}
\caption{(Color online) Final Monte Carlo (MC) magnetization $M_f$ as a function of the scaled sweep rate $v$ for
various values of $w$. Each data point is an average of 100 runs for a system of $N=82{,}519$ spins using the original
transition rate (\ref{gamma1}). Also shown is \eno{m_f_k} for $M_{f,{\rm K}}$, the final magnetization in the
Kayanuma model, and the implied final magnetization from the experimental data for \Fe8~\cite{wer99}.}
\label{m_f-v}
\end{figure}
As can also be seen in \fno{m-bias}, the final magnetization, $M_f$, agrees with the Kayanuma result only for large
enough $v$ (the agreement with the LZS result is even poorer). To make this point clearer, in \fno{m_f-v} we show
$M_f$ vs.~the sweep rate $v$ for three different choices of $w$, along with the Kayanuma result
\beq
	M_{f,{\rm K}} = 1 - 2 P_{\rm K} = e^{-\pi\Dta^2/|\dot\eps_a|} = e^{-\pi/v}. \label{m_f_k}
\eeq
There is good agreement with the Kayanuma result for fast sweep rates, with $v \gtwid 10$. In fact for these
velocities, the Kayanuma and LZS models both give good answers. However, we do not compare the Monte Carlo answers with
LZS because the disagreement between them for $v \ltwid 10$ is much worse. Below $v$ of about $10$, even the agreement
with Kayanuma's model starts to degrade, and we obtain a residual magnetization even for very slow sweep rates. The
agreement is worse for smaller values of $w$, with a lower value of $w$ corresponding to a larger $M_f$. This is
qualitatively understandable as larger $w$ allows for greater opportunities for relaxation. What is interesting is
that for $v \gtwid 10$, there is virtually no $w$ dependence in $M_f$, and the divergence in $M_f$ starts to appear
at about the same value of the scaled velocity, $v \sim 10$, as that where LZS and Kayanuma start to disagree.

\begin{figure}[h!]
\centering
\includegraphics[scale=.8]{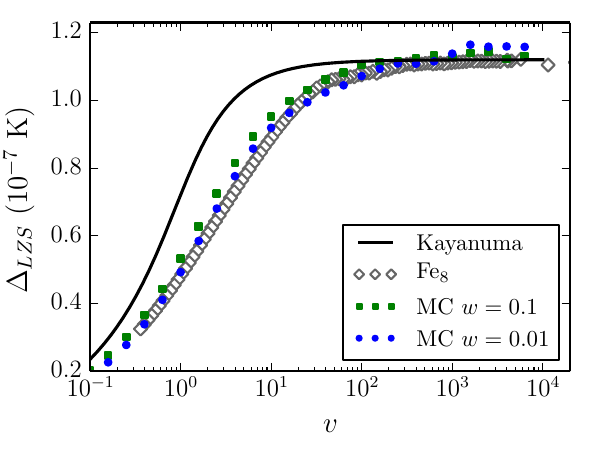}
\caption{(Color online) The value of $\Dta$ that we would infer for \Fe8 if we applied the LZS formula for $M_f$ to the
Kayanuma model, our Monte Carlo (MC) simulations, and the experimental data from Wernsdorfer and Sessoli.~\cite{wer99} (open
diamonds). The MC simulation parameters are the same as in \fno{m_f-v}. Since the actual dependence of $M_f$ on $v$ is incorrectly
given by LZS, fitting to LZS leads to a sweep-rate dependent answer for the inferred $\Dta$. For the experimental data, we deduce the
conversion factor from $\dot\eps$ to $dH/dt$ by exploiting the fact that for high sweep rates, LZS and Kayanuma both give the
correct answer for $\Dta$, viz. $1.12 \times 10^{-7}\,$K.}
\label{delta_inf}
\end{figure}
We now show what value of $\Dta$ we would infer if, as experimenters often do, we use the LZS formula for the final
magnetization~\cite{AM}. In other words, denoting this inferred value by $\Dta_{{\rm inf,LZS}}$, we use the formula
\beq
\Dta^2_{{\rm inf,LZS}} = - \frac{2 |{\dot\eps}|}{\pi} \ln \Bigl( \frac{1 + M_f}{2} \Bigr).
\eeq
The results are shown in \fno{delta_inf}. We show what we get if we use the Kayanuma formula for $M_f$, the
experimental data of Ref.~\cite{wer99}, and $M_f$ as given by our Monte Carlo simulations with $w = 0.01$ and
$w=0.1$. From this plot, one cannot tell which value of $w$ is better, although as \fno{m_f-v} shows, $w = 0.01$
gives a better fit. Since we estimated $W \simeq 10\,$mK and $E_{dm} \simeq 100\,$mK, $w = 0.1$ certainly seems
reasonable, and even $w = 0.01$ is not out of the question. This graph shows that the magnetization reversal is
not a very sensitive function of $w$, and so is of limited value in trying to learn about nuclear spin decoherence.
As can be seen, the disagreement between even the Kayanuma model and the experimental data is substantial (observe
the logarithmic scale for $v$), while that between the data and our simulations is small.
%

%
\begin{figure}[h!]
$\begin{array}{cc}
 \includegraphics[scale=.60]{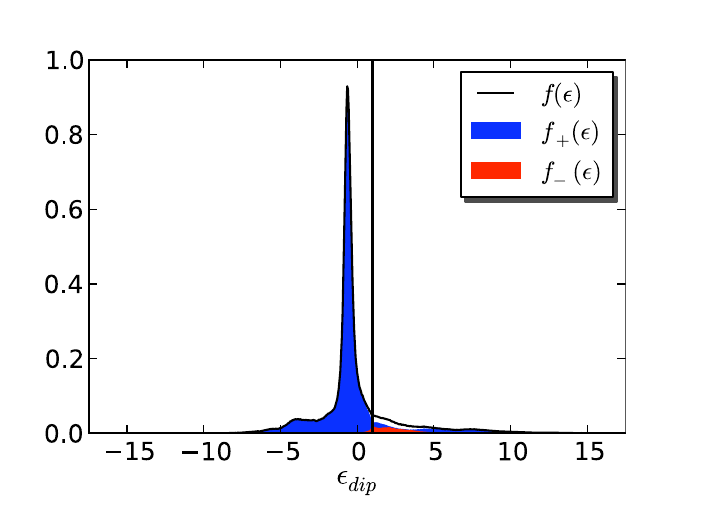} &
 \includegraphics[scale=.60]{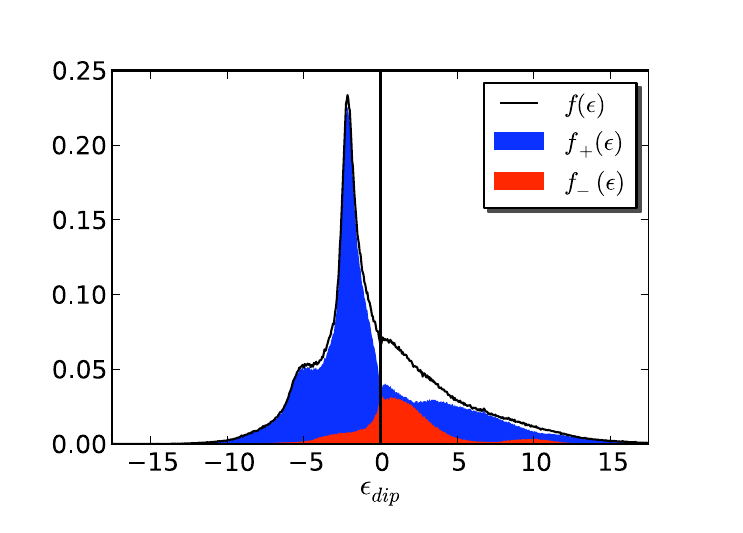} \\
 \includegraphics[scale=.60]{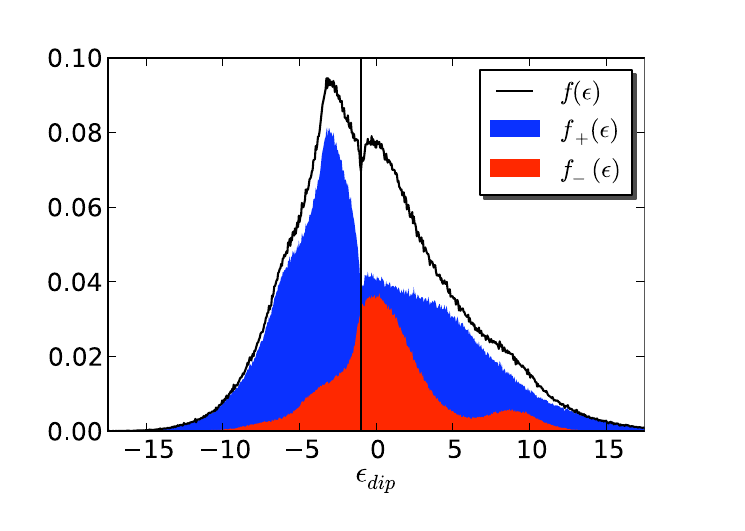} &
 \includegraphics[scale=.60]{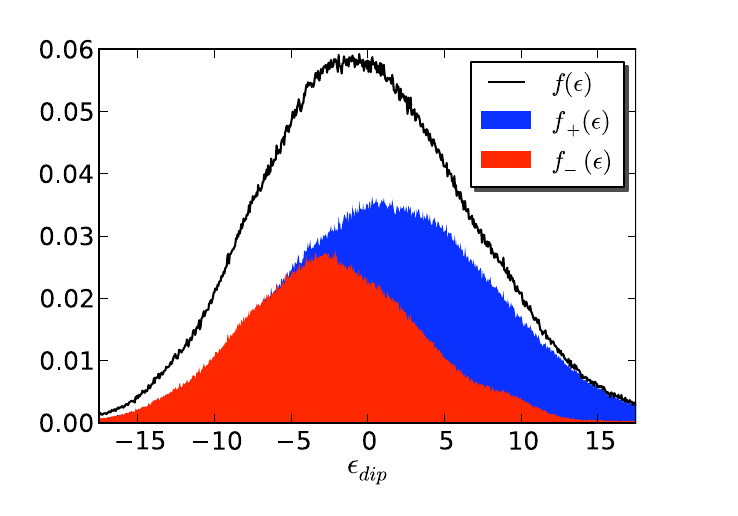}
\end{array}$
\caption{(Color online) Time dependence of the bias distributions $f_{\pm}(\veps)$ and
$f(\veps) = \sum_{\sig} f_{\sig}(\veps)$
over the course of a field sweep for a sample of $N=20{,}479$ spins with $v = 0.1$, $w = 0.05$ at (a) $\eps_a=-1.0E_{dm}$,
(b) $\eps_a=0.0E_{dm}$, (c) $\eps_a=1.0E_{dm}$, (d) $\eps_a=25.0E_{dm}$, all averaged over 100 runs of the Monte Carlo
simulation with the modified rate function (\ref{gamma2}). The reversible region of the bias distribution is very
narrow on the scale of this figure, and shows up as the vertical line.}
\label{biascomp}
\end{figure}
We now attempt to understand the nonzero residual final magnetization. To this end, we consider the bias
distributions $f_{\pm}(\eps)$ over the course of a field sweep. Initially, the bias distribution at up spin sites
is sharply peaked at zero, with only spins on the surface of the sample having a significant nonzero bias. As the external
bias $\eps_a$ approaches zero, spins begin to flip, causing the distribution to widen. Even though the dipolar
interaction is long-ranged, when a spin flips from up to down, the resulting change in bias on the majority of the
sample is small on the scale of $W$. However, spins within a couple of lattice spacings of the spin that flipped
will experience a dramatic change in bias. Of the six nearest neighbors of a flipping spin, the four in the \itxy
plane will experience a change in bias of $-4E_{dm}$, and the two along the \itx axis will experience a change
$8E_{dm}$. This can be seen in \fno{biascomp}b as the bias distribution has two minor peaks displaced approximately
$-4E_{dm}$ and $8E_{dm}$ from the central one. Suppose that we are at a point in the sweep where a particular test
spin in the sample is seeing a net bias close to zero, and one of its nearby spins (not necessarily a nearest
neighbor) flips. We refer to the second spin as the triggering spin. Let the first spin undergo a large negative
change in bias as a result of this spin flip event. It will then be displaced in bias far from the reversible region,
yet since the external bias is being swept from negative to positive values, it will reenter the reversible region
at a later time provided that the dipolar contribution to the bias seen by it has not changed significantly. On the
other hand, if the test spin undergoes a large positive change in bias, it will be displaced in bias far past the
reversible region. Unless the triggering spin flips yet again or another nearby spin flips so as to change the bias
on the test spin by a large negative amount, it will continue to see a large positive bias, and remains in its
original orientation, up. It will essentially have been shunted around the region of reversibility. These shunted
spins give a net positive contribution to $M_f$, while unshunted spins will contribute an amount that is close to
the Kayanuma result on average.

\begin{figure}[h!]
\centering
\includegraphics[scale=.6]{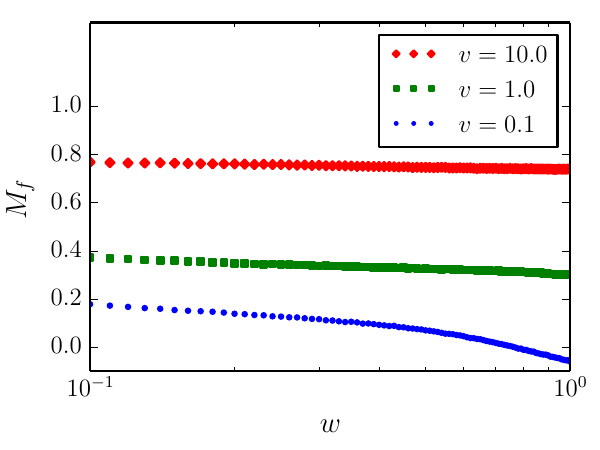}
\caption{(Color online) Final Monte Carlo magnetization as a function of the reversible region width $w$ for several
small values of $v$, using the Gaussian rate function (\ref{gamma1}). Each data point is an average of 100 runs for
a system of $N =7{,}153$ spins.}
\label{m_f-w}
\end{figure}
We conclude this subsection by discussing two atypical simulations. First, we investigate the dependence of $M_f$
on $w$. We ran the simulation for fixed $v \le 10$ and variable $w$ using the original Gaussian rate function
(\ref{gamma1}). The results are plotted in \fno{m_f-w} and show that the $w$ dependence is weak, although it is
more pronounced for smaller $v$ and $M_f$ tends to the Kayanuma result with increasing $w$.

\begin{figure}[h!]
$\begin{array}{cc}
	\includegraphics[scale=.60]{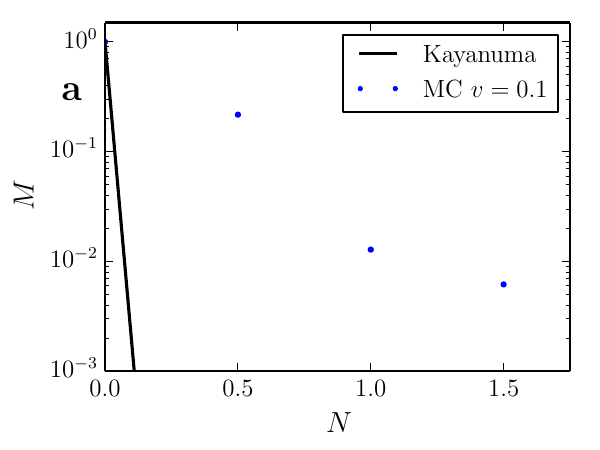} &
	\includegraphics[scale=.60]{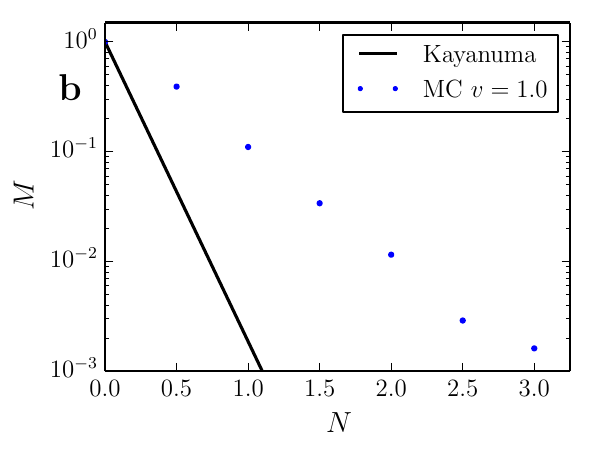} \\
	\includegraphics[scale=.60]{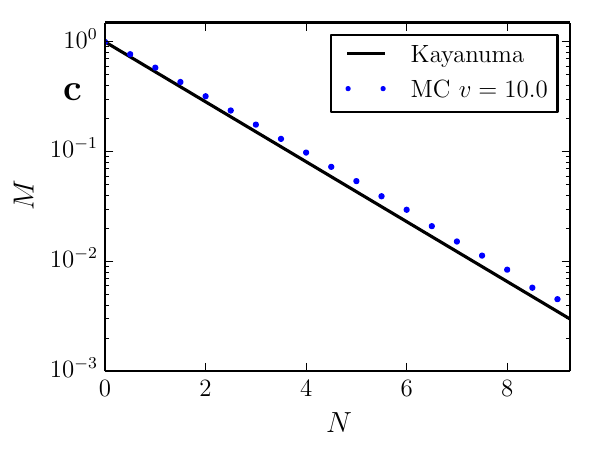} &
	\includegraphics[scale=.60]{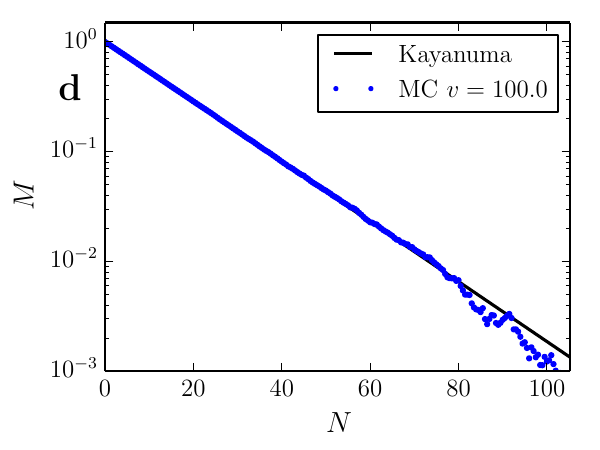}
\end{array}$
\caption{(Color online) Monte Carlo results for the magnetization $M$ plotted against number of cycles $N$ for an
applied triangular wave using the original Gaussian flip rate (\eno{gamma1}) with $w=0.05$ and $v=0.1$, 1, 10,
and 100. The grey line shows the prediction of the Kayanuma model. The simulations are performed with 7{,}153 spins.
In the figure for $v = 100.0$, the scatter in the data at large $N$ is statistical.}
\label{M_triang}
\end{figure}
Second, we show the results of simulations in which $\eps(t)$ is a triangular wave. In \fno{M_triang} we show the
dependence of $M$ on the number of cycles of the wave for different values of $v$. Note that we show $M$ every
half-cycle, where a cycle is a complete period of the wave. The relevant point here is that $M$ decreases
essentially exponentially with the number of cycles, except for very small $v$, showing that it is the fraction
$\Dta M/M_i$ that is the same each time the reversible region is traversed, where $M_i$ is the initial fraction
before the traversal. This is true even when the agreement with the Kayanuma model is poor.

\section{Kinetic Equation}
\label{kin_eqn}

In Ref.~\cite{vij12}, the results of Monte Carlo simulations of magnetization relaxation were analyzed in terms of
coupled rate equations for three quantities: the magnetization $M$, the magnetization of the spins in the reversible
bias region $M_r$, and the number of spins in the reversible region $N_r$. However, the equations did not form a
closed system, and also involved a functional $\mathcal{F}$, given by
\beq
\mathcal{F}[f_{\sig}(\eps)]
    = W^2 \sum_{\sig} \int_{\left| \eps \right|>W} \frac{f_{\sig}(\eps)}{\eps^2}\, d\eps.
\eeq
The distribution $\sum_{\sig} f_{\sig} (\eps)$ was then handled via an approach called the Three Gaussian
Approximation (TGA), wherein it was modeled as a sum of three Gaussians, centered at biases of $0$, corresponding
to a spin with no neighbors flipped, and $-4E_{dm}$ and $8E_{dm}$, corresponding to spins with one nearest neighbor
flipped. The widths of the Gaussians were assumed to be equal, and this common width and the heights of the three
Gaussians were determined by matching the first three moments of $\sum_{\sig} f_{\sig}(\eps)$ with a model in which
every spin was up or down with probabilities $(1 \pm M)/2$ independently of the others. The distribution determined
in this way depends only on the system's magnetization, and leads to a closed system of rate equations. These
equations were then found to describe the short to moderate time behavior of the magnetization well.

It is plain that this approach is inadequate for the swept-field problem. It is paramount to have a good
approximation for the bias distribution near the reversible region to properly capture the probability of spin
flips. For the relaxation problem, the reversible region is static and centered at zero bias, whereas in the case
of swept field, the reversible region moves over the full range of the bias distribution. As a result, a larger
number of spins are capable of flipping, and the bias distribution is altered over essentially it entire range.
Thus, the TGA is fundamentally invalid and one cannot really identify just three peaks. The approximation may perhaps
work for very fast sweeps since the fraction of spins that flip is then small, but even in this limited success it
does not provide a physically satisfactory explanation.

It is therefore necessary to analyze the time evolution of the full bias distribution $f_{\sig}(\eps)$. This is
naturally done in terms of a kinetic equation. As stated in \Sno{intro}, this can be obtained from that given in
Ref.~\cite{pro98ab} by replacing the two-site bias distribution with a product of single-site distributions. In
addition, we must allow for an explicit change in the applied field because of the sweep. Rather than present the
equation as merely the outcome of this formal procedure, however, we develop it by examining the microscopic
processes that change the bias distribution and give rise to spin flips, as this gives a better appreciation of
the physics. The basic processes are exactly the same as in Ref.~\cite{vij12}, so our discussion here is more brief.

\subsection{Derivation of kinetic equation}
\label{derive_ke}

Let us denote the rate at which a spin at bias $\eps$ flips from $\sig$ to $-\sig$ by
\beq
\gsbs(\eps),
\eeq
where we write $\bar\sig$ for $-\sig$ in the suffixes. In general the rates $\gssb(\eps)$ and $\gsbs(\eps)$ need not
be equal, although in this paper they are. One way in which $\fsi(\eps)$ can change is by what might be called
direct or one-spin processes, wherein in a time $dt$, a spin flips at any given site $i$, but does not undergo any
change in the bias seen by it. This gives a contribution
\beq
   \lf. \frac{d\fsi}{dt}\rt|_{\rm direct} = - \gsbs(\eps) \fsi(\eps) + \gssb(\eps) \fsb(\eps).  \label{df_direct}
\eeq

Next, we consider two-spin processes, where we focus on a spin at a particular site, $i$, and allow the bias seen
by this spin to change by virtue of a spin-flip at another site $j$. We refer to the second spin as the
{\it triggering\/} spin, and to the first as the {\it central\/} spin. The latter is taken to not flip, as the
probability for a process where it also flips is of order $(dt)^2$ and therefore negligible. There are then two
processes which cause $\fsi(\eps)$ to change:
\beq
\begin{array}{rccccc}
          &       {\rm site\ }i     &         {\rm site\ }j         &     &       {\rm site\ }i        &    {\rm \ site\ }j            \\
{\rm I.}  &  \quad\sig, \eps\quad   &    \quad\sig', \eps'\quad   & \to &    \quad\sig, \eps''\quad   &  \quad-\sig', \eps'\quad  \\
{\rm II.} &  \quad\sig, \eps''\quad  &    \quad\sig', \eps'\quad   & \to &    \quad\sig, \eps\quad    &  \quad-\sig', \eps'\quad
\end{array}
\eeq
Processes I and II can be thought of as loss and gain processes, respectively, since they lead to spins being knocked
out of or into the bias region $(\eps, \eps + d\eps)$. Let us consider process I first. Since the spin at site $j$
flips from $\sig'$ to $-\sig'$, the change in this spin is $-2\sig'$. In order for the bias at site $i$ to change as
shown, we must have
\beq
\eps'' = \eps - 2 K_{ij} \sig', \label{esp_change}
\eeq
i.e.,
\beq
K_{ij} = \tshf \bigl(\eps - \eps'') \sig'.
\eeq
If we let the final bias on site $i$ lie in the range ($\eps'', \eps'' + d\eps''$), the number of triggering spins,
i.e., the number of spins that satisfy this condition is
\beq
g\bigl(\tshf(\eps - \eps'')\sig' \bigr) \times \frac{d\eps''}{2},
\eeq
where
\beq
g(K) \equiv \sum_{j \ne i} \dta(K-K_{ij}) \label{def_gk}
\eeq
is the density of couplings, by which we mean that $g(K) dK$ is the number of sites which couple to the central site
with couplings between $K$ and $K + dK$. The sum in \eno{def_gk} is over an infinite lattice, and therefore
independent of site $i$. The probability that a spin on a triggering site will indeed flip in time $dt$ is
$\gspp\,dt$. To find the net loss in $\fsi(\eps)$, we must multiply the number of triggering sites with the
probability of a flip at those sites and the fraction of sites $i$ and $j$ that have the stipulated spins and
biases. We must then sum over all possible values of $\sig'$, $\eps'$, and $\eps''$. In this way we get
\beq
\lf. \frac{d\fsi}{dt}\rt|_{\rm I}
    = - \sum_{\sig'}\int d\eps' \int \frac{d\eps''}{2}
            \gspp \geepp f_{\sig'}(\eps') \fsi(\eps). \label{df_I}
\eeq

The calculation for process II proceeds in identical fashion. This time we get
\beq
\lf. \frac{d\fsi}{dt}\rt|_{\rm II}
    = \sum_{\sig'}\int d\eps' \int \frac{d\eps''}{2}
            \gspp \geppe f_{\sig'}(\eps') \fsi(\eps''). \label{df_II}
\eeq
Together, the last two equations describe what might be called the collision integral.

Lastly, let us incorporate a swept or explicitly time dependent applied field. If there were no spin flip
processes at all, then a site which at time $t$ had a bias $\eps$, would at time $t + dt$ have a bias
\beq
\eps(t+dt) = \eps(t) + d\eps_a,
\eeq
where $d\eps_a$ is the change in the applied field in the interval $dt$, and we would have
\bea
\fsi(\eps(t), t)
    &=& \fsi\bigl(\eps(t) + d\eps_a, t+dt \bigr) \nnu\\
    &=& \fsi\bigl(\eps(t), t \bigr) + \frac{\ptl\fsi}{\ptl\eps} d\eps_a + \part{\fsi}{t} dt.
\eea
Therefore,
\beq
\lf. \frac{d\fsi}{dt}\rt|_{\rm sweep} = - \part{\fsi}{\eps} \frac{d\eps_a}{dt}. \label{df_sweep}
\eeq

Adding Eqns.~(\ref{df_direct}), (\ref{df_I}), (\ref{df_II}), and (\ref{df_sweep}), we get
\bea
\frac{d\fsi}{dt}
    &=& - \part{\fsi}{\eps} \frac{d\eps_a}{dt}
        - \gsbs(\eps) \fsi(\eps) + \gssb(\eps) \fsb(\eps) \nnu\\
       && - \sum_{\sig'}
           \dint \frac{d\eps'\,d\eps''}{2} 
             f_{\sig'}(\eps') \gspp \Bigl[ \geepp \fsi(\eps)  - \geppe \fsi(\eps'') \Bigr]. \nnu\\
       &&{\hskip5pt}
   \label{df_tot}
\eea
This is the kinetic equation we are seeking. We note once again that it is also the equation we would get from
that in Ref.~\cite{pro98ab} upon factorization of the two-site bias distribution function, and addition of the
explicit time dependence of $\eps_a$. We can rewrite this equation in several other ways which help understand
its structure, and also suggest algorithmic simplifications for numerical integration. So, we note that in the
collision term, the integral over $\eps'$ can be factored out of the expression completely, reflecting the fact
that the {\it precise value of the bias\/} at the triggering site is irrelevant to whether a flip on this site
alters the bias on the central site by a given amount; what matters for that is {\it where\/} the triggering site
is located relative to the central site, and the spin on the triggering site. We may therefore define a {\it net
rate per site\/} for spin flip, averaged over the bias distribution, and over the entire sample,
\beq
\tsbs(t) = \int d\eps\, f_{\sig}(\eps, t) \gsbs(\eps).
\eeq
As indicated, this rate changes with time because the bias distribution changes. It is in fact a functional of
the distribution. In terms of this rate, we may write the kinetic equation as
\bea
\frac{d\fsi}{dt}
    &=& - \part{\fsi}{\eps} \frac{d\eps_a}{dt}
        - \gsbs(\eps) \fsi(\eps) + \gssb(\eps) \fsb(\eps) \nnu\\
       && - \hf\sum_{\sig'} \tspp
                  \int d\eps'\, \Bigl[ \geep \fsi(\eps)  - \gepe \fsi(\eps') \Bigr].
   \label{df_tot2}
\eea
We have also changed the dummy variable of integration from $\eps''$ to $\eps'$.

An elementary but useful check on the kinetic equation is that the quantity
\beq
\sum_{\sig}\int d\eps\, \fsi(\eps)
\eeq
should be time-independent, since it is just unity by normalization. Summing \eno{df_tot2} over $\sig$ and
integrating over all $\eps$, we get
\bea
\frac{d}{dt}\lf[\sum_{\sig} \int d\eps\, \fsi(\eps) \rt]
    &=& - \frac{d\eps_a}{dt} \lf[\sum_{\sig} \int d\eps\, \part{\fsi}{\eps} \rt]
        - \sum_{\sig} \int d\eps\, \Bigl[ \gsbs(\eps) \fsi(\eps) - \gssb(\eps) \fsb(\eps) \Bigr] \nnu\\
          && - \hf\sum_{\sig, \sig'} \tspp
                  \dint d\eps\,d\eps'\, \Bigl[ \geep \fsi(\eps)  - \gepe \fsi(\eps') \Bigr]. \nnu\\
     &&{\hskip5pt}
   \label{d_norm}
\eea
The sweep term integrates to zero directly, since $\fsi(\eps)$ must vanish for $\eps \to \pm\infty$. The direct terms
also add to zero. To see that, we first note that a sum over $\sig$ is equivalent to a sum over $\bar\sig$. Recognizing
this, and writing the second direct term as a sum over $\bar\sig$, and then interchanging the index labels $\sig$ and
$\bar\sig$, the two direct terms are seen to cancel each other identically. To see that the two-spin terms also add
to zero, we simply interchange the integration variables $\eps$ and $\eps'$ in the very last term. The two parts of
the collision integral are then identical and they cancel each other. Hence,
\beq
\frac{d}{dt}\lf[\sum_{\sig} \int d\eps\, \fsi(\eps) \rt] = 0,
\eeq
as desired.

At this point it is useful to discuss the nature of $g(K)$, the density of dipole couplings. As shown in
Ref.~\cite{avag09}, 
\beq
	g(K) \apx \frac{16 \pi}{9 \sqrt{3}}\frac{E_{dm}}{K^2}, \quad K \to 0\pm.  \label{gk_asymptotic}
\eeq
That is, for small $K$, $g(K)$ is approximately an even function of $K$, and diverges as $1/K^2$. The integrand of the
$\eps'$ integral in \eno{df_tot2} therefore has a singularity at $\eps' = \eps$ of the form
\beq
\frac{\fsi(\eps') - \fsi(\eps)}{(\eps' - \eps)^2},
\eeq
which is equivalent to
\beq
\frac{1}{\eps' - \eps} \part{\fsi(\eps)}{\eps}.
\eeq
This is an integrable singularity if we view the integral over $\eps'$ as a principal value. Hence, our kinetic
equation is mathematically well posed. It is also clear that this singularity cannot have any physical consequences.
It arises from flips of triggering spins which are very far away from the central spin, and these flips cannot
affect the dynamical behavior of the central spin since they lead to miniscule changes in the bias seen by the
latter. Fortunately, it is not necessary to take any special precautions about this singularity in the numerical
integration, since it is automatically regulated by the finiteness of the sample.

We can understand the last point further as follows. In the collision terms, the integrals over $\eps'$ are
equivalent to a sum over sites. Consider, for example, the two-spin loss term (process I). We can write
\beq
\lf. \frac{d\fsi}{dt}\rt|_{\rm I}
       = - \fsi(\eps) \sum_{\sig'} \tspp
                  \int \frac{d\eps'}{2}\, \geep   \label{df_loss_1}
\eeq
Now,
\bea
\int \frac{d\eps'}{2}\, \geep 
    &=& \sum_{j\ne i} \int d\eps'\, \dta\bigl(\eps' - (\eps - 2K_{ij})\sig'\bigr) \nnu\\
    &=& \sum_{j\ne i} 1. \label{df_loss_2}
\eea
In the same way, for the gain term (process II), we have
\bea
\int \frac{d\eps'}{2} g\bigl(\tshf(\eps' - \eps)\sig'\bigr) \fsi(\eps')
  &=& \sum_{j \ne i} \int \frac{d\eps'}{2} \dta\bigl(\tshf(\eps' - \eps)\sig' - K_{ij}\bigr) \fsi(\eps') \nnu\\
  &=& \sum_{j \ne i} \fsi(\eps + 2 \sig' K_{ij}).
\eea
Hence the combined contribution of these two processes can be written as
\beq
\lf. \frac{d\fsi}{dt}\rt|_{\rm I + II}
       =  \sum_{\sig'} \tspp \sum_{j\ne i} \bigl( \fsi(\eps + 2\sig' K_{ij}) - \fsi(\eps) \bigr).
\eeq
Let us suppose that we cut off the sum so that site $j$ lies inside a large sphere (centered at site $i$) containing
$N_K$ sites. For spins outside this sphere, $K_{ij}$ is very small in magnitude, and we may approximate
\beq
\sum_{j > N_K} \bigl( \fsi(\eps + 2\sig' K_{ij}) - \fsi(\eps) \bigr)
    = \sum_{j > N_k} 2 \part{\fsi(\eps)}{\eps} \sig' K_{ij},
\eeq
which vanishes if we divide the sum into subsums carried out over sets of sites related by cubic symmetry.

We also note that the form (\ref{gk_asymptotic}) for $g(K)$ is a poor estimate for near neighbor or short-distance
couplings. It is these couplings which are in the end responsible for the peaked form of the bias distribution
function, and must therefore be treated correctly taking the discontinuous delta-function nature into account. In
other words, we must handle the part of the integral where $\eps' - \eps$ is large as a sum over neighboring sites,
and a continuum form for $g(K)$ cannot be used.

\subsection{Numerical Integration}

To integrate the kinetic equations, the bias distribution functions $f_+(\eps)$ and $f_-(\eps)$ were approximated
by histograms with a scaled bias range $[-\veps_{\rm max},\veps_{\rm max}]$, divided into $N_b$ bins. As in the Monte
Carlo simulations, $\veps_{\rm max}$ was set to 25. At first sight it appears that we should choose the bin width,
$w_b = 2\veps_{\rm max}/N_b$, to be much less than $w$. Since realistic values of $w$ are quite small, however, this
would require the number of bins, $N_b$ to be rather large. Now a larger number of bins leads to a smaller number of
spins in each bin, and since we want the {\it relative\/} change in this number per time step to be small for accurate
integration, it requires smaller integration timesteps. Hence too small a bin width is numerically expensive. In
practice we find that changing $w_b$ only affects how very small changes in bias are handled, and it is perfectly
acceptable to let the bin width equal the reversible region width $2w$, leading to $N_b$ in the range
$10^3$--$10^4$. Smaller values of $w_b$ did not lead to appreciably different results.

Since the kernel $K_{ij}$ is a property solely of the lattice, we determine it once and for all before doing any
integration as soon as we have decided upon the bin width $w_b$. All values of $-K_{ij}$ are determined for a large
sphere of $N_K$ spins with the site $i$ at the center. The quantity $-K_{ij}$ is half the amount by which the bias
will change at the central site if a spin $\sig_j$ flips from up to down. We therefore compile a histogram from this
data with a bin width matching that of the bias distribution functions, giving us
$g\bigl((\eps-\eps')/2 \bigr) d\eps$.
It should be noted that it does no good to bin $g(K)$ more finely. Let us denote this bin width by $w_g$. The value
in a given bin of the $g(K)$ histogram essentially gives us the number of sites that can shift the bias by $2K$.
The value of $g(K)$ in the next bin will count the sites that shift the bias by $2(K+w_g)$. If $2w_g$ is smaller
than $w_b$, we are needlessly differentiating between sites that have the same effect as far the evolution of the
histogram for $\fsi(\eps)$ is concerned. We therefore choose $2w_g$ equal to $w_b$.

Combining \etwo{df_loss_1}{df_loss_2} for the contribution to $df/dt$ from process I (the loss term), we get
\beq
\lf. \frac{d\fsi}{dt}\rt|_{\rm I} = - N_K (T_{{\bar 1}1} + T_{1{\bar 1}}) \fsi(\eps)
\eeq
We can combine this with the term from process II to write
\beq
\lf. \frac{d\fsi}{dt}\rt|_{\rm I + II}
       =  \hf\sum_{\sig'} \tspp \int d\eps'\, \gpepe \fsi(\eps'), \label{df_coll}
\eeq
where
\beq
g_R(K) = g(K) - N_K \dta(K) \label{def_greg}
\eeq
is a regulated density of couplings.

The form (\ref{df_coll}) once again shows that the infrared divergence mentioned in the previous subsection is a
numerical nonissue. Once we have settled on a bin width $w_b/2$ for the histogram of $g(K)$, there is no point in
increasing $N_K$ beyond a certain value. Beyond that value, we only change the number in the central bin around
$K=0$. The delta-function term in \eno{def_greg} goes into the same bin, so $g_R(K)$ does not change. Thus the
numerics are essentially done for an infinite system, and the width of the bin serves as a cutoff that effects
the principal value integral.

It also pays to define the population {\it transfer\/} rate
\beq
\Gam(\eps, \eps') = \hf\sum_{\sig'} \tspp \gpepe. \label{def_Gam_matrix}
\eeq
When we multiply this quantity by the bin width $w_b$, the quantity $\tshf \gpepe w_b$ is the number of sites
capable of triggering a shift in bias from $\eps'$ to $\eps$, and $\tspp$ is the spin-flip rate averaged over all
sites, so $w_b \Gam(\eps, \eps')$ is the rate at which population is transferred from the bin containing the bias
$\eps'$ to the bin containing $\eps$. It also gives the entire collision integral a nifty matrix multiplication form,
\beq
\lf. \frac{d\fsi}{dt}\rt|_{\rm I + II}
       =  \int d\eps'\, \Gam(\eps,\eps') \fsi(\eps'), \label{df_coll_matrix}
\eeq
and the complete kinetic equation can be written as
\beq
\frac{d\fsi}{dt}
    = - \part{\fsi}{\eps} \frac{d\eps_a}{dt}
        - \gsbs(\eps) \fsi(\eps) + \gssb(\eps) \fsb(\eps) \nnu\\
          +  \int d\eps'\, \Gam(\eps,\eps') \fsi(\eps').
   \label{df_tot3}
\eeq

The actual integration of the kinetic equations is simple once the histograms for $f_+$, $f_-$, and $g_R(K)$ are
set up. In each timestep, there is some exchange of the populations in the central bins of the $f_+$ and $f_-$
histograms, corresponding to the reversible region, as dictated by \eno{df_direct}. Next, the collision terms are
evaluated for every bin in the two histograms, given by \eno{df_coll}, and the integration moves to the next timestep.
The reversible region is swept from $-\eps_{\rm max}$ to $\eps_{\rm max}$ for LZS runs, while it is allowed to remain
static at the origin for relaxation runs. 

\subsection{Kinetic Equation Results}

As a test of the kinetic equations and our numerical integration procedure, we first applied them to the problem of
magnetic relaxation with zero external bias. The equations were integrated up to a time $10\tau$ where $\tau$ is the
characteristic time for relaxation, given by
\beq
\tau = \frac{E_{dm}}{\pi\Dta^2}.
\eeq
\begin{figure}[h!]
\includegraphics[scale=.8]{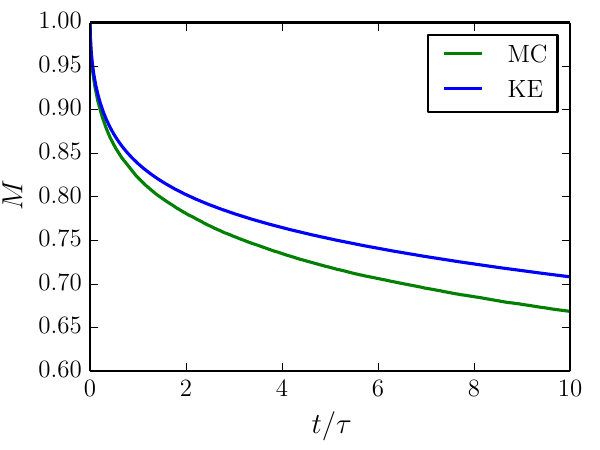}
\caption{(Color online) Comparison of Monte Carlo (MC) and kinetic equation (KE) results for magnetic relaxation
with $w = 0.05$. MC data was obtained using $N=82{,}519$, and averaged over 20 runs. KE data was obtained using
2{,}001 bins and a scaled bias range $(-50,50)$. Note that the zero on the vertical scale is suppressed.}
\label{demagke}
\end{figure}
(Note that this time scale is very long in an absolute sense since it varies as $\Dta^{-2}$, and $\Dta$, being a
tunnel splitting between deep levels on opposite sides of the anisotropy barrier, is very small.) The resulting
demagnetization curve is plotted in Fig. \ref{demagke} along with the results of the Monte Carlo simulation. As can
be seen, it agrees well with the Monte Carlo simulation for short times. However, after $t/\tau \gtwid 1$,
the two curves begin to diverge, with the kinetic equations giving a higher value for the magnetization than the
Monte Carlo simulation. This difference can be understood by comparing the bias distributions given by the two
methods. These distributions are shown in \fno{bias_relax}.

\begin{figure}[h]
$\begin{array}{cc}
 \includegraphics[scale=.80]{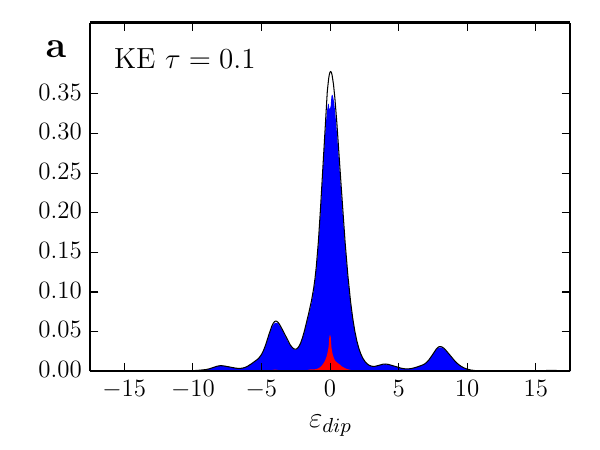} &
 \includegraphics[scale=.80]{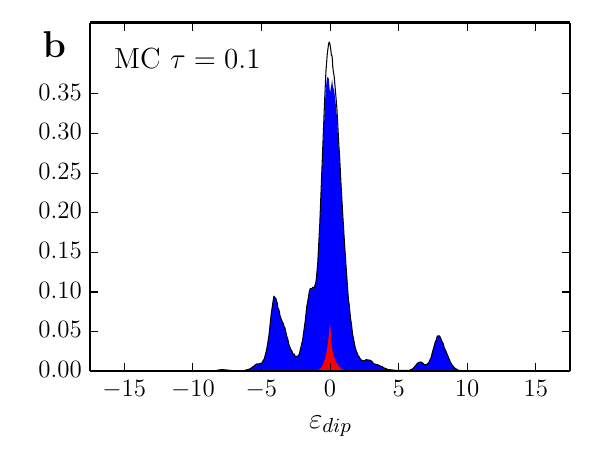} \\
 \includegraphics[scale=.80]{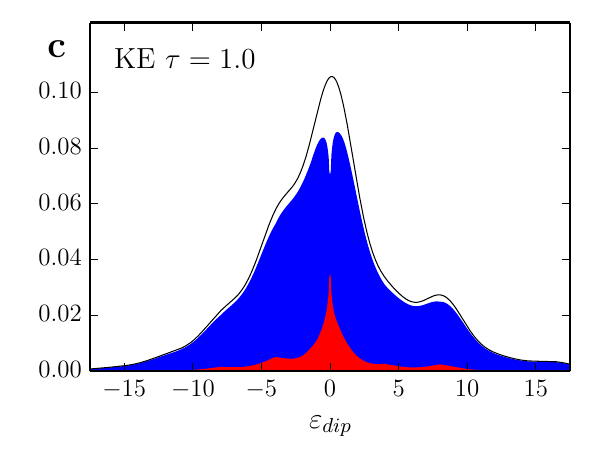} &
 \includegraphics[scale=.80]{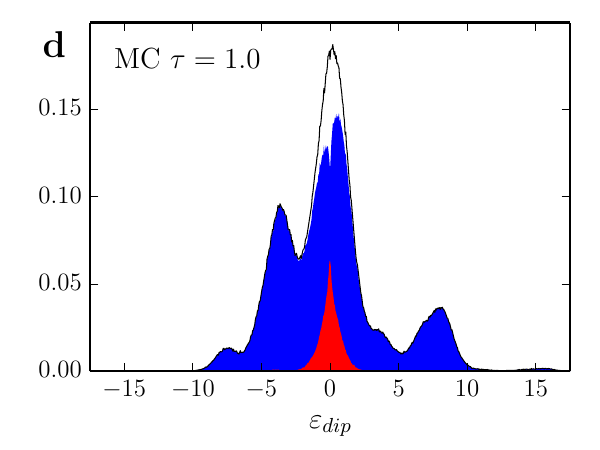} \\
\end{array}$
\caption{(Color online) Comparison of how the bias distributions evolve during magnetic relaxation as found by
solving the kinetic equation (KE) (left panels) and by Monte Carlo (MC) simulations (right panels). The scaled time
$t/\tau$ is $0.1$ in the top two figures, and $1.0$ in the bottom two. The (scaled) reversibility region width
$w = 0.05$ in all figures. MC data were obtained using $N=82{,}519$ and averaged over 20 runs. KE data were
obtained using 2{,}001 bins and a scaled bias range $(-50,50)$. The curve labels and colors are the same as in
Fig.~\ref{biascomp}.}
\label{bias_relax}
\end{figure}

For short times, $t \ltwid 0.1\tau$, the bias distributions match fairly well, with each peak in the Monte Carlo
data also present in the solution to the kinetic equations. However, after $t \simeq \tau$, the bias distribution
from the Monte Carlo is bounded by about $\pm 10 E_{dm}$, and still shows sharp side peaks, while that from the
kinetic equations is significantly more spread out and lacking the peaks. This difference is due to the fact that
the kinetic equations do not carry information about the orientations and biases of a spin's neighbors while the
Monte Carlo simulation does. More specifically, there is a short distance correlation between biases, as we now discuss.
Consider a central spin and its 6 nearest neighbors all initially in the up state and with zero bias. If one of the
neighboring spins flips, the bias on the central spin will change by an amount $8E_{dm}$ or $-4E_{dm}$, moving it out
of the reversible region to one of the peaks that can be seen in Fig. \ref{bias_relax}b. Similarly, the bias at
the five other neighboring sites will also change and these spins will also be removed from the reversible region.
As the sample continues to relax, the bias for the central spin will not have a large probability to change
significantly because of the fact that five of its nearest neighbors are displaced far from the reversible region.
The sixth spin, the one that originally flipped, is still sitting in its original near-zero bias, and therefore has
a higher probability to flip back, but this would cause the bias for the central spin to move back closer to the
origin. Thus, by considering a spin's neighbors, we can see that once a spin acquires a large bias, its neighbors
are more likely than not to also acquire a large (but perhaps not quite as large) bias. This condition of the
neighboring spins then results in a low probability for the bias on the central spin to change yet again by a large
amount, and if such a change does occur it is more likely to make the bias regress back toward the mean. This
explains why the bias distribution in the Monte Carlo simulation is more bounded and peaked. The kinetic equations
do not keep track of these correlations, and thus yield too broad a distribution.

We then applied the kinetic equations to the problem of a swept field. In this case, the equations are much more
successful, and, as shown in \fno{m_f_ke}, we find that the results are in good agreement with the Monte Carlo
simulation for values of the scaled sweep rate, $v$, as low as 5, where the Kayanuma model does quite poorly.
However, the limitations of the kinetic equations again show up for $v \ltwid 5$.

\begin{figure}[h]
\includegraphics[scale=.8]{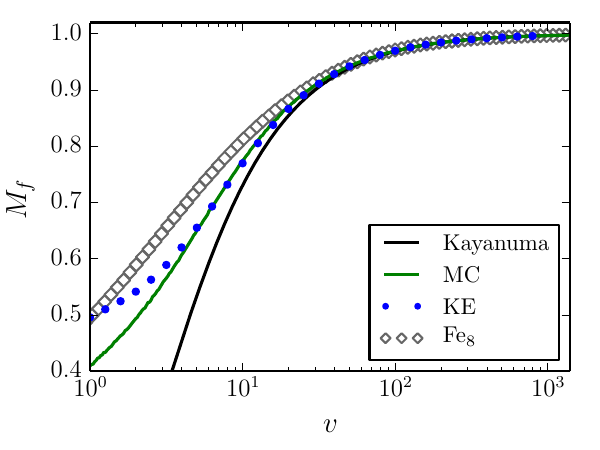}
\caption{(Color online) $M_f$ vs{.} $v$ as given by the Kayanuma model, our Monte Carlo simulations (with a width
parameter $w = 0.05$), and the kinetic equation (with $w = 0.05$, and a bin width $E_{dm}/30$). Also shown are the data
for \Fe8 from Ref.~\onlinecite{wer99}}
\label{m_f_ke}
\end{figure}

\section{Conclusions}
\label{conc}

We have carried out Monte Carlo simulations of swept-field experiments on molecular magnetic solids based on the
microscopic view of spin reversal developed in Refs.~\cite{avag09,pro96}. We find that these simulations provide a
very good picture of the time evolution of the entire system, and agree fairy well with experiments quantitatively.
In order to understand the simulations, we have also developed a kinetic equation for the distribution of
single-site spin and bias distribution. This kinetic equation also provides a quantitatively accurate description
of experimental data even for quite low sweep velocities. However, the kinetic equation fails at very low sweep
velocities, since it is then incapable of accounting for important short-distance bias correlations. Expanding the
kinetic equation approach to include two-site distributions is non trivial and difficult. Nevertheless, the kinetic
equation should be capable of describing many more experimental protocols, and we hope to do this in the future.

\acknowledgments
We are indebted to Nandini Trivedi for useful discussions and for very generously allowing us the use of her
computer cluster at Ohio State University for some of the numerical work.

\newpage
\voffset=1.0in
\hoffset=0.5in
%
%

\begin{thebibliography}{99}
\bibitem{fe8props}
For a discussion of the many experiments that allow these and other properties of \Fe8 to be deduced, see
Ref.~\cite{gat06}, Sec.~4.7.4, and references therein.
\bibitem{bias} It is apparent that the bias $\eps_i = 2\mu H_i$, where
$\mu$ is the magnetic moment of a molecule, and $H_i$ is the magnetic
field at the $i$th site. We prefer to work with the bias as all
interactions in the problem are then expressed in terms of energies.
\end{thebibliography}
\end{document}